\documentclass[10pt]{article}

\usepackage{amsmath}
\usepackage{amssymb}
\usepackage{cite}
\usepackage{hyperref}
\usepackage{lineno}
\usepackage[margin=3cm]{geometry}

\newif\iffigmod
\newsavebox\figcaps
\ifdefined\ebcdcchallsetupfile
  \input\ebcdcchallsetupfile
\else
  \figmodfalse
  \newenvironment{condpreview}{}{}
  \newenvironment{condfigure}[1][]{\begin{figure}[#1]}{\end{figure}}
\fi
\newcommand\shortcite\cite
\newcommand\s{s}
\newcommand\tests{{s_{\mathrm{curr}}}}

\newcommand\histbase{y}
\newcommand\hist[1]{{\histbase^{#1}}}

\newcommand\underlyingfbase{f}
\newcommand\underlyingtaubase{\tau}
\newcommand\underlyingf[1]{{\underlyingfbase^{#1}}}
\newcommand\underlyingtau[1]{{\underlyingtaubase^{#1}}}
\newcommand\fitfbase{{\hat{f}}}
\newcommand\fitf[1]{{\fitfbase^{#1}}}
\newcommand\fitfdraw{f}
\newcommand\fitfdrawxform[1]{{f^{#1}}}
\newcommand\fittaubase{{\hat{\tau}}}
\newcommand\fittau[1]{{\fittaubase^{#1}}}
\newcommand\eps\epsilon

\newcommand\uchoice{\mathop{\mathsf{Unif}}}
\newcommand\tarj{{\mathrm{tar}_j}}

\newcommand\cvtests{{s_{\mathrm{cv}}}}
\newcommand\mult[1]{{\mathrm{#1}}}

\DeclareMathOperator*{\argmax}{arg\,max}
\DeclareMathOperator*{\avg}{avg}

\usepackage{tikz}                                                           %%%%
\usetikzlibrary{shapes,positioning,arrows,backgrounds}                      %%%%
\usepackage{float}                                                          %%%%
\usepackage{subcaption}                                                     %%%%
\title{Flexible Modeling of Epidemics \\ with an Empirical Bayes Framework}
\author{
Logan~C.~Brooks$^1$
\and
David~C.~Farrow$^1$
\and
Sangwon~Hyun$^2$
\and
Ryan~J.~Tibshirani$^2$
\and
Roni~Rosenfeld$^1$\thanks{\texttt{Roni.Rosenfeld@cs.cmu.edu}}
}
\date{
$^1$\emph{School of Computer Science, Carnegie Mellon University, \\
Pittsburgh, Pennsylvania, United States of America}
\\
$^2$\emph{Department of Statistics, Carnegie Mellon University, \\
Pittsburgh, Pennsylvania, United States of America}
}

\begin{document}

\maketitle

\section{Abstract}
Seasonal influenza epidemics cause consistent, considerable, widespread loss annually
in terms of economic burden, morbidity, and mortality.
With access to accurate and reliable forecasts of a current or upcoming influenza epidemic's behavior,
policy makers can design and implement more effective countermeasures.
This past year, the Centers for Disease Control and Prevention
hosted the ``Predict the Influenza Season Challenge'',
with the task of predicting key epidemiological measures
for the 2013--2014 U.S. influenza season
with the help of digital surveillance data.
We developed a framework for in-season forecasts of epidemics
using a semiparametric Empirical Bayes framework,
and applied it to predict
the weekly percentage of outpatient doctors visits for influenza-like illness,
as well as the season onset, duration, peak time, and peak height,
with and without additional data from Google Flu Trends.
Previous work on epidemic modeling has focused on
developing mechanistic models of disease behavior and
applying time series tools
to explain historical data.
However, these models may not accurately capture the range
of possible behaviors that we may see in the future.
Our approach instead produces possibilities for the epidemic curve of the season of interest
using modified versions of data from previous seasons,
allowing for reasonable variations in the timing, pace, and intensity of the seasonal epidemics,
as well as noise in observations.
Since the framework does not make strict domain-specific assumptions,
it can easily be applied to other diseases as well.
Another important advantage of this method is that
it produces a complete posterior distribution for any desired forecasting target,
rather than mere point predictions.
We report prospective influenza-like-illness forecasts
that were made for the 2013--2014 U.S. influenza season,
and compare the framework's cross-validated prediction error on historical data
to that of a variety of simpler baseline predictors.
% ~293--298 words
% todo sum up results

% Please keep the Author Summary between 150 and 200 words
% Use first person. PLOS ONE authors please skip this step. 
% Author Summary not valid for PLOS ONE submissions.   
\section{Author Summary}
Influenza epidemics occur annually,
and incur significant losses in terms of lost productivity, sickness, and death.
Policy makers employ countermeasures, such as vaccination campaigns,
to combat the occurrence and spread of infectious diseases,
but epidemics exhibit a wide range of behavior, which makes
designing and planning these efforts difficult.
Accurate and reliable numerical forecasts of how an epidemic will behave,
as well as advance notice of key events,
could enable policy makers to further specialize countermeasures
for a particular season.
While a large amount of work already exists on modeling epidemics in past seasons,
work on forecasting is relatively sparse.
Specially tailored models for historical data may be overly strict
and fail to produce behavior similar to the current season.
We designed a framework for predicting epidemics
without making strong assumptions about how the disease propagates
by relying on slightly modified versions of past epidemics
to form possibilities for the current season.
We report forecasts generated for the 2013--2014 influenza season,
and assess its accuracy retrospectively.
% ~163-167 words
% todo mention CDC challenge, sum up results

\section{Introduction}
Seasonal influenza epidemics occur each year
and incur significant economic burden, morbidity, and mortality.
The annual impact in the United States
has been estimated at
$611\mult{K}$ lost undiscounted life-years,
$3.1\mult{M}$ hospitalized days,
$31.4\mult{M}$ outpatient visits, and
$\$87.1\mult{B}$ in economic burden~\cite{molinari2007annual}.
Accurate and reliable forecasts
offer many opportunities to improve
preparedness and response to influenza epidemics.
Long-term predictions could be used
to help select a vaccine
for the next season.
Forecasts within a season
can help
policy makers to tailor vaccination campaigns and advisories,
hospitals to prepare staff and beds,
and individuals and organizations to plan for vaccination and potential sickness.
Despite the notable impacts of the disease, though,
many weaknesses of influenza surveillance and prediction systems in the past~\cite{Laporte1873}
remain today.
Capabilities to observe and forecast the prevalence of influenza and similar diseases
lag considerably, e.g., behind analogues in meteorology.
During the 2013--2014 flu season,
the CDC hosted the ``Predict the Influenza Season Challenge'',
which encouraged teams to forecast features of the current epidemic progression that would be useful to policy makers,
and to take advantage of digital surveillance such as search engine and social network data.
The competition established a closer relationship
between forecasters and policy makers,
and provided valuable assessment of the performance of true (prospective) within-season forecasts.

Existing work on modeling influenza epidemic curves
generally falls into one of three categories:
\begin{description}
\item[Compartmental models] estimate the number of people
in various states related to a disease~\cite{hethcote2000mathematics}.
For example, the SIR model approximates dynamics
between the proportions of the population
susceptible to influenza,
infected with the virus,
and recovered from infection.
Common assumptions include that any pair of individuals in a population are equally likely to interact,
and that different strains of influenza behave identically.
% todo citations
\item[Agent-based models] generate synthetic populations based on census data
and build complex schemes of interaction and disease behavior in synthetic humans~\cite{ferguson2006strategies,colizza2007modeling,bansal2006comparative,lee2010simulating,grefenstette2013fred}.
% todo citations from FRED paper (petal/pedal?)
It is common for these systems to be applied to
the special case of a single, novel strain of influenza.
\item[Parametric statistical models] are tools from time series modeling
that are less closely tied with mechanistic assumptions of how flu is transmitted.
Simple approaches include linear autoregression,
which estimates flu activity at some time with a linear function of the flu activity in the recent past.
More complex alternatives include Box-Jenkins analysis, generalized linear models (GLM), and generalized autoregressive moving-average models~\cite{Dugas2013}.
\end{description}

% todo reorder... influenza matters, policy makers now, predictions helpful; predictions lag, cdc challenge, ...?

Past forecasting efforts~\cite{Chretien2014,Nsoesie2014a} usually take a
compartmental model~\cite{Shaman2012a,Ong2010},
agent-based model,
or parametric statistical model~\cite{Goldstein2011,Vergu2006,soebiyanto2010modeling},
and condition on partial data
to predict flu activity levels one to ten weeks in the future.
Other methods include
prediction markets~\cite{Polgreen2007},
which combine expert predictions using a stock market-like system,
and the method of analogues ($k$ nearest neighbors)~\cite{viboud2003prediction},
which makes predictions of future flu activity levels
using similar patterns from the past,
without assuming a strict model.
% todo include Nsoesie Dirichlet process model?  simulated data though
% todo more/less citations? other competitors?

We take a nonmechanistic approach,
generating possibilities for 
the current season's epidemic curve
using modified versions of past seasons' curves,
incorporating reasonable adjustments in the timing, pace, and intensity of the epidemic,
as well as accommodating noise in observations.
Our method differs from the mechanistic and parametric statistical model approaches
in that it models the process generating the data nonparametrically,
rather than using a model that may significantly misrepresent the data.
While the method of analogues is similar in this regard as a nonparametric method,
our framework considers the entire season as a unit,
which differs from the traditional perspective in nearest neighbor modeling.
Our framework also models the error in observations,
and outputs a distribution rather than point predictions,
while existing applications of the method of analogues
generate single point predictions one at a time.
% todo last sentence in this paragraph
% todo some summing up of results?

\section{Materials and Methods}
\subsection{Surveillance data}
% todo add links on how to obtain
\subsubsection{U.S. Outpatient Influenza-like Illness Surveillance Network (ILINet)}
The Centers for Disease Control and Prevention (CDC) release several forms of surveillance data
% xxx "release" vs "releases"?  if spelled out, "release"; if abbreviated, "releases"; here, ???
regarding the prevalence, type, and impact of influenza-like illness (ILI)
in the United States~\cite{brammer2013seasonal,cdc2013overview}.
These data (as well as GFT and our predictions) are in terms of ILI,
because doctors do not generally diagnose influenza specifically,
but rather as part of a broader syndromic category of ILI.
Since ILI is generally not notifiable,
its activity is measured not with case counts,
but with the percentage of doctor's visits that are ILI-related
during a given epidemiological week.
The U.S. Outpatient Influenza-like Illness Surveillance Network (ILINet)
is a group of over $2,900$ outpatient healthcare providers that
voluntarily provide information about
the number of total visits and ILI-related visits that they receive.
The CDC compiles ILINet reports,
adjusts for effects of changes in participation,
and weights data based on state population.
The result, called percent weighted ILI (wILI),
is released on a weekly basis, with about a weekly delay for reporting and processing,
at a national level and for each of the ten Health \& Human Services (HHS) regions,
broken down by age group;
data may be revised in later weeks.
% xxx note that revisions were most noticeable in last three weeks?
This data is available for every season since the 1997--1998 season.\footnote{%
% The CDC did not report wILI data for weeks 21--39 in earlier seasons.
% Beginning with the 2003--2004 season, wILI data is reported for every week.
The CDC did not report wILI data for weeks 21--39 in the first six seasons of ILINet surveillance.
Beginning with the 2003--2004 season, wILI data is reported for every week.
}
%xxx mention unavailability in earlier seasons in "off-season"?

\begin{condfigure}
  \centering
  \begin{condpreview}
  \begin{tikzpicture}
    \newlength\aspc
    \setlength\aspc{1.5em}
    \tikzstyle{latent} = [draw, ellipse, fill=black!15]
    \tikzstyle{observedish} = [draw, ellipse, densely dashed, fill=black!15]
    \tikzstyle{observed} = [draw, ellipse]
    \tikzstyle{deterministic} = [->,ultra thick,>=stealth]
    \tikzstyle{stochasticpartial} = [->,>=stealth]

    \node[latent, align=center] (influenza_incidence) {Influenza \\ incidence};
    \node[latent, right=\aspc of influenza_incidence, align=center] (other_ili_incidence) {Other ILI \\ incidence};
    \path (influenza_incidence) -- (other_ili_incidence)
      node[midway, below=1.4\aspc, latent] (ili_incidence) {ILI incidence};
    \node[latent, below=\aspc of ili_incidence] (ili_visits) {ILI visits};
    \node[observedish, below=\aspc of ili_visits] (ilinet_ili_visits) {ILINet ILI visits};
    \node[observed, below=\aspc of ilinet_ili_visits] (wILI) {wILI};

    \node[latent, right=2\aspc of ili_visits, align=center] (ili_queries) {Google queries tied \\ to ILI incidence};
    \node[latent, right=\aspc of ili_queries, align=center] (other_queries) {Other Google \\ queries};
    \path (ili_queries.south) -- (other_queries.south)
      coordinate[midway] (querieshor) {};
    \node[observedish, align=center] (queries) at (querieshor |- ilinet_ili_visits) {Google queries};
    % \node[observed, right=5em of wILI, align=center] (past_wili) {Past (w)ILI};
    \node[observed, below=\aspc of queries, align=center] (GFT) {GFT};

    \draw[deterministic] (influenza_incidence) to (ili_incidence);
    \draw[deterministic] (other_ili_incidence) to (ili_incidence);
    \draw[stochasticpartial] (ili_incidence) to (ili_visits);
    \draw[stochasticpartial] (ili_visits) to (ilinet_ili_visits);
    \draw[deterministic] (ilinet_ili_visits) to (wILI);
    \draw[stochasticpartial] (ili_incidence) to (ili_queries);
    \draw[stochasticpartial] (ili_visits) to (ili_queries);
    \draw[deterministic] (ili_queries) to (queries);
    \draw[deterministic] (other_queries) to (queries);
    \draw[deterministic] (queries) to (GFT);
    % \draw[deterministic] (past_wili) to (GFT);
    % Begin from http://tex.stackexchange.com/questions/64505/how-to-make-each-pdf-page-of-a-beamer-output-have-an-opaque-background-when-it-i.
    \begin{pgfonlayer}{background}
      \fill[white]
      (current bounding box.south east)
      rectangle
      (current bounding box.north west);   
    \end{pgfonlayer}
    % End from http://tex.stackexchange.com/questions/64505/how-to-make-each-pdf-page-of-a-beamer-output-have-an-opaque-background-when-it-i.
  \end{tikzpicture}
  \end{condpreview}
  
  \caption{%
Generation process for ILINet and GFT data.
We are interested in influenza and other ILI incidence,
but cannot observe them directly.
Instead, we rely on wILI as a measure of flu prevalence,
and sometimes use GFT to approximate wILI.
Shaded nodes, unobserved quantities;
shaded dashed nodes, proprietary data;
unshaded nodes, publicly available data;
thin arrows, dependencies;
thick arrows, deterministic dependencies.
  }
  \label{fig:data_net}
\end{condfigure}

\subsubsection{Google Flu Trends (GFT)}
Google Flu Trends (GFT) is a system designed to
estimate (``nowcast'') CDC ILINet data
up to and including the current week
using Google query data.
GFT results are available in near real-time,
with final estimates of ILI activity in a given week
available soon after that week ends.
% todo where are the partial-week GFT numbers?  when are the final ones for a week available?
Estimates are available for
the nation as whole and the ten HHS regions,
as well as smaller geographical units such as states.
The original algorithm~\cite{Ginsberg2009}, launched in 2008,
was updated in 2009~\cite{Cook2011} and 2013~\cite{Copeland2013}
to improve performance by regenerating its selection of queries using additional data,
and by revising the method itself.
Despite these modifications,
GFT has recently drawn criticism~\cite{Lazer2014,santillana2014can}
on a number of issues, including its performance versus some simple alternatives.
% xxx maybe cite lazer2014google?  (follow-up, not peer-reviewed?)
However, existing work at the start of the competition
indicated that GFT was the most accurate of
existing digital surveillance systems~\cite{lamb2013separating},
and is helpful when used in combination with CDC ILINet data~\cite{Dugas2013}.
% xxx Lamb2011 citation has wrong type or missing fields in Epi Forecasting collection
We used GFT results as a proxy for CDC ILINet data for a few weeks before our predictions were made,
when CDC data was not yet released, or could be revised significantly later.
\autoref{fig:data_net} illustrates the relationship between ILINet data, GFT data, and underlying phenomena.

\subsection{Empirical Bayes framework}
The forecasting framework is composed of five major procedures:
\begin{enumerate}
\item Model past seasons' epidemic curves as smoothed versions plus noise.
\item Construct prior for the current season's epidemic curve by considering reasonable sets of transformations of past seasons' curves.
\item Estimate what the wILI values in recent past will be after their final revisions, using non-final wILI and GFT.
\item Weight possibilities for current season's epidemic curve using estimates of final revised wILI.
\item Calculate forecasting targets for each possibility, and report results.
\end{enumerate}

The first two steps only need to be executed once,
at the beginning of the current season.
As additional data becomes available throughout the season,
we generate forecasts using steps 3--5.

We perform predictions for each geographical unit
--- the U.S. as a whole or individual HHS regions ---
separately.
Historically, surveillance has focused on influenza activity
between epidemiological weeks $40$ and $20$, inclusive.
% xxx lcb: apparently flu season is "defined" as this period, but I cannot find anything online to that effect
We define seasons as epidemic weeks $21$ to $39$, the ``preseason'',
together with weeks $40$ to $20$.
During the competition, data was available for $15$ historical seasonal influenza epidemics.
We excluded the 2009--2010 season from the data
since it included nonseasonal behavior from the 2008--2009 pandemic in the preseason.
Additionally, there was partial data available for the 2013--2014 season.

\subsubsection{Data model}
We view wILI trajectories as the sum of
some underlying ILI curve and plus noise:
\begin{equation}
\hist\s(i) = \underlyingf\s(i) + \eps_i^\s,
\qquad \eps_i^\s\sim\mathcal{N}(0, \underlyingtau\s),
\qquad \text{for each week }i,
\label{eqn:data_model}
\end{equation}
where
$\hist\s(i)$ is the wILI value for the $i$th week of season $\s$,
$\underlyingf\s$ is the underlying curve, and
$\eps_i^\s$ is normally distributed noise.
We estimate the underlying ILI curve $\fitf\s$
from the wILI curve $\hist\s$
with quadratic trend filtering~\cite{tibshirani2014adaptive}
for each historical season $\s$.
This method smooths out fluctuations in the wILI data,
producing a new set of points that lie on a piecewise quadratic curve.\footnote{
  The quadratic trend filtering procedure produces one point for each available wILI observation,
  i.e., 33 or 34 for the first six seasons, and 52 or 53 for the rest.
  We fill in the curve on the rest of the real line by
  copying the first available wILI value at earlier times,
  copying the last measurement at later times,
  and using linear interpolation at non-integer values.
  These filled-in values are later used by the peak week and pacing transformations.
}
We estimate the level of noise using the one-standard-deviation rule:
% xxx how is this the one-standard-deviation rule?
\[(\fittau\s)^2=\avg_i[\hist\s(i)-\fitf\s(i)]^2.\]

\subsubsection{Prior}
The key assumption of the framework is that the current season
will resemble one of the past seasons,
perhaps with a few changes.
\begin{description}
\item[Shape:] 
% xxx notes about SIR stuff?
The general shape $\fitfdraw$ of the underlying curve is
taken from one of the past seasons.
We select each of the historical shapes with equal probability:
$\fitfdraw\sim\uchoice\{\fitf\s:\text{historical season }\s\}$.
% $\uchoice\trainss$ to determine which $\esthist\s$ to use.
% \trains a bit confusing since capital version does not correspond to pdist
\item[Noise:]
The standard deviation of the normally distributed noise at each week
is assumed to take on values from the past years' candidates with equal probability:
$\sigma\sim\uchoice\{\fittau\s:\text{historical season }\s\}$.
\item[Peak height:]
The distribution of underlying peak heights is drawn from 
a continuous uniform distribution: $\theta\sim U[\theta_m,\theta_M]$.
We use an unbiased estimator for $\theta_m$ and $\theta_M$
based on past seasons' trend filtered curves.
\item[Peak week:]
The distribution of underlying peak weeks is formed
in a similar manner to the peak height distribution;
we find unbiased estimators $\mu_m$, $\mu_M$ for
%continuous
uniform distribution bounds,
but restrict the distribution to integral output:
$\mu\sim\uchoice\{i\in\{1..53\}: \mu_m\le i\le \mu_M\}$.
\item[Pacing:]
We allow for variations in the ``pace'' of an epidemic
by incorporating a time scale
that stretches the curve about the peak week;
the distribution of time scale factors is $\nu\sim U[0.75,1.25]$.
% xxx simplify, since there's an equation anyway?
\end{description}
To generate a possible curve for the current season,
i.e., to sample from the prior,
we independently sample a shape, noise level, peak height, peak week, and pacing parameter
from the above distributions,
then generate the corresponding wILI curve.\footnote{%
We have also developed and are investigating
an alternative ``local'' transformation prior
that does not use information from other historical curves
when transforming a particular historical curve $f$,
but instead reuses the noise level for $f$
and makes smaller \emph{changes} to the peak week and height of $f$,
which are restricted to a smaller, predefined range.
}

We model the underlying curve $\underlyingf\tests$ for the current season
as the curve generated by
a randomly sampled parameter configuration
$\left\langle\fitfdraw,\sigma,\nu,\theta,\mu\right\rangle$,
using the following equation:
\[\fitfdrawxform\tests(i)=b+\frac{\theta-b}{\max_j\fitfdraw(j)-b}\left[\fitfdraw\left(\frac{i-\mu}\nu+\argmax_j\fitfdraw(j)\right)-b\right],\]
% todo check
where $b$ is the current year's baseline wILI level (i.e., the onset threshold)
for the selected geographical region, e.g., $2\%$ for the U.S. as a nation for the 2013--2014 flu season.
\autoref{fig:peakxforms} illustrates the peak week and peak height transformations.
% xxx should introduce the onset threshold, explain what it is, cite
\begin{condfigure}
  \centering
  \begin{condpreview}
  \begin{subfigure}{0.49\linewidth}
    \includegraphics[width=\linewidth]{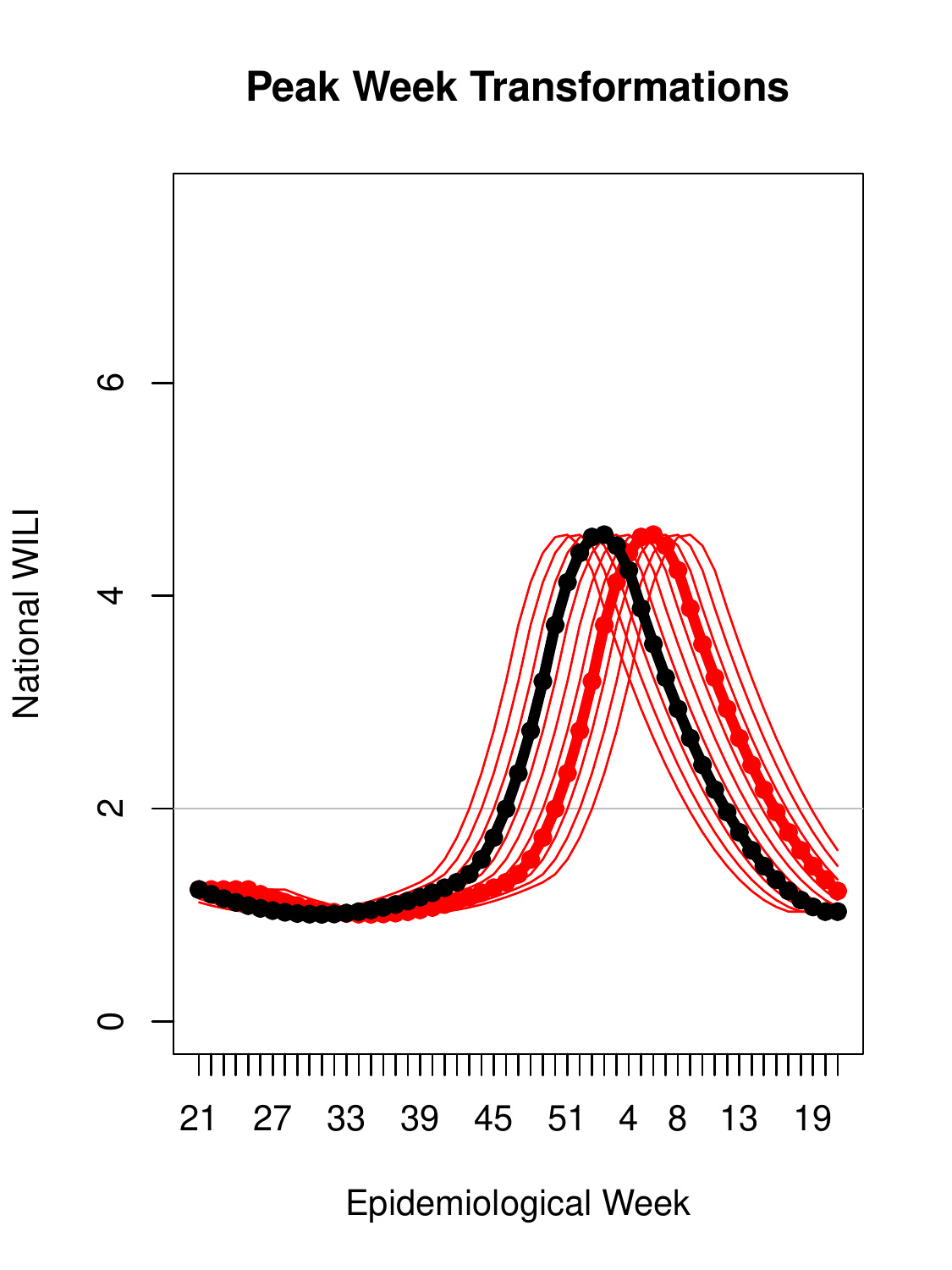}
    % \vspace{-2\baselineskip}
    % \caption{Peak week transformations.  Black, original curve; red, possible curves after peak week transformation.}
    \caption{}
    \label{fig:peakweakxforms}
  \end{subfigure}
  \begin{subfigure}{0.49\linewidth}
    \includegraphics[width=\linewidth]{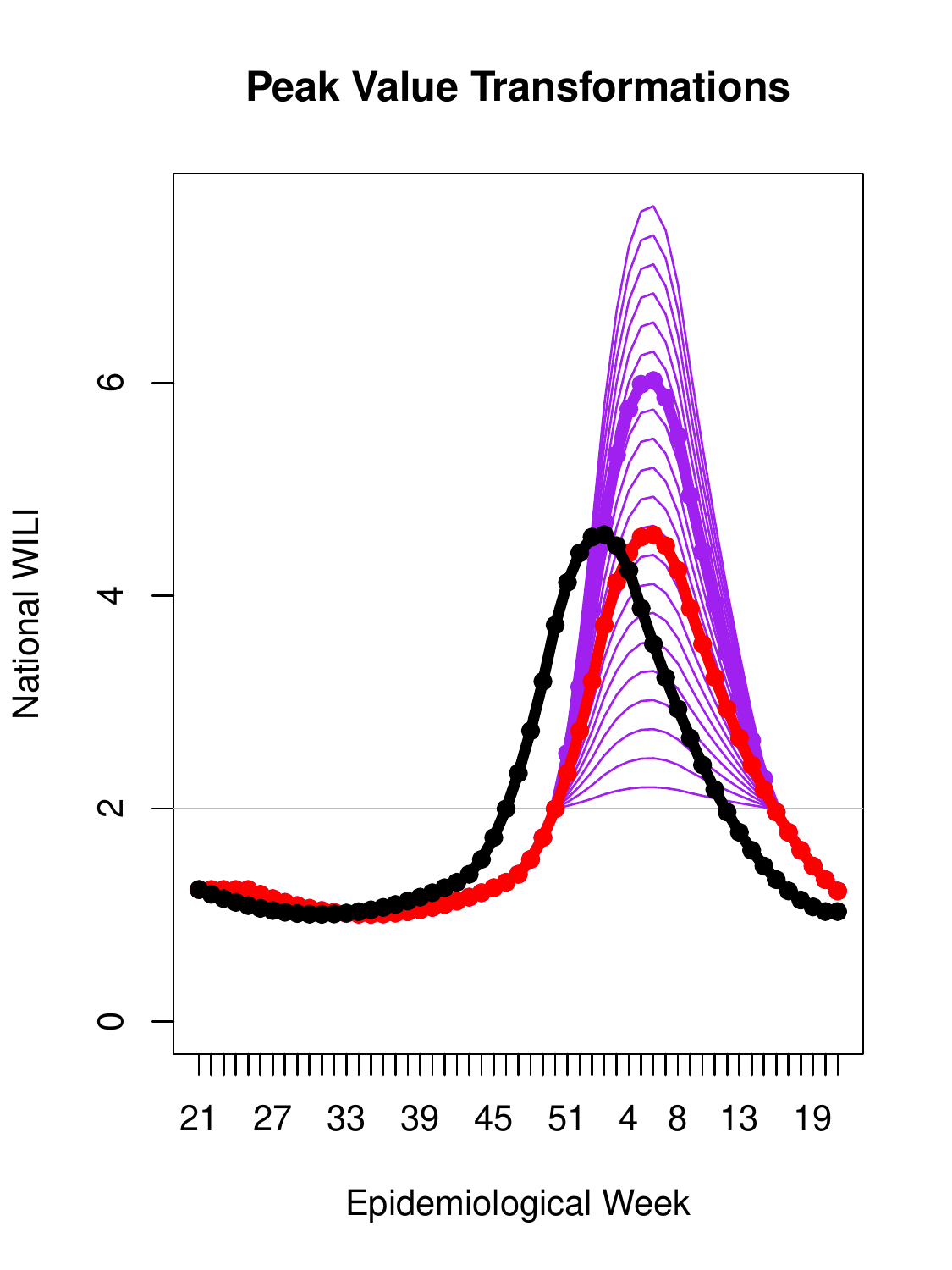}
    % \vspace{-2\baselineskip}
    % \caption{Peak height transformations.  Black, original curve; red, curve after peak week transformation; purple, possible curves after peak height transformation.}
    \caption{}
    \label{fig:peakheightxforms}
  \end{subfigure}
  \end{condpreview}
  \caption{%
Examples of possible peak week and peak height transformations.
Thick black, original curve;
red, possible peak week transformations;
thick red, a random peak week transformation;
purple, possible peak height transformations;
thick purple, a random peak height transformation.
(a) Peak week transformations.
Peak weeks of historical smoothed curves occurred between weeks 51 and week 10 of the next year,
so we limit transformations to give peak weeks roughly within this range.
(b) Peak height transformations.
Peak heights of historical smoothed curves were between 2\% and 8\%,
so we limit transformations to give peak heights roughly within this range.
}
  \label{fig:peakxforms}
\end{condfigure}
The data model for the current season's wILI values $\hist\tests$
is the same as that for historical seasons,
shown in \autoref{eqn:data_model}.

\subsubsection{Sampling from the posterior}
We use importance sampling~\cite{liu2008monte}
% todo reference supplementary material
to obtain a large set of curves from the posterior
weighted by how closely they match the epidemic curve
so far, beginning with week 40.
More concretely, we obtain a single weighted sample from the posterior by
(i) sampling a historical smoothed curve $f$, noise level $\sigma$, and transformation parameters $\nu$, $\theta$, and $\mu$ from the prior;
(ii) applying the peak height, peak week, and pacing transformations;
(iii) assigning the curve an ``importance weight'' or ``likelihood'' based on how well it matches existing observations for the current seasons; and
(iv) drawing noisy wILI observations around the curve for the rest of the season.
We apply this procedure many times to obtain a collection of
possible wILI trajectories and associated weights,
forming a probability distribution over possible futures for the current season.

\subsubsection{Forecasting targets}
For the CDC challenge,
we were interested in four forecasting targets
of interest to policy makers:
the epidemic's onset, peak week, peak, and duration.
\begin{description}
\item[Onset:] The first week that the wILI curve
is above a specified CDC baseline wILI level,
and remains there for at least the next two weeks.
For example, the 2013--2014 national baseline wILI level was $2\%$,
so the onset was the first in at least three consecutive weeks
with wILI levels above $2\%$.
\item[Peak Week:] The week in which the wILI curve
attains its maximum value.
\item[Peak:] The maximum observed wILI value in a season.
\item[Duration:] Roughly, how many weeks the wILI level remained
above the CDC baseline since the onset.
We defined this more rigorously as the sum of the lengths
of all periods of three or more consecutive weeks
with wILI levels above the CDC baseline.
\end{description}
We generate distributions for each of these targets by repeatedly
(i) sampling a possible wILI trajectory and associated weight from the posterior,
(ii) calculating the four forecasting targets for that trajectory,\footnote{
When calculating the forecasting targets for a particular wILI trajectory,
we use the observed wILI values when they are available.
This ensures that the framework will not consider
target values that seem impossible with the currently available data,
e.g., peak height values lower than the currently observed maximum.
Future data revisions may make some of these previously ``impossible'' values valid again, though.
} and
(iii) storing these four values along with the trajectory's weight.
We represent these forecasting target posterior distributions with histograms, and
generate point estimates by taking the posterior mean for each target.
\autoref{fig:framework_net} illustrates the links between the elements of the framework.

\begin{condfigure}
\centering
\begin{condpreview}
\begin{tikzpicture}
  \tikzstyle{latent} = [draw, ellipse, fill=black!15]
  \tikzstyle{observed} = [draw, ellipse]
  % xxx change arrows to use styles from above
  % todo add intermediate $\fitf\tests$ between params and $\esthist\tests$
  % \node[latent] (TrainFit) {Historical $\fitfbase$, $\fittaubase$};
  \node[latent] (TrainFit) {Historical $\underlyingfbase$, $\underlyingtaubase$};
  \node[observed, above=of TrainFit] (TrainDat) {Historical $\histbase$};
  \node[latent, below=of TrainFit] (Hyper) {$\theta_m$, $\theta_M$, $\mu_m$, $\mu_M$};
  \node[latent, right=of TrainFit] (Params) {$\fitfdraw$, $\sigma$, $\theta$, $\mu$, $\nu$};
  \node[right=2em of Params] (Data) {};
  \node[observed] (Obs) at (Data |- TrainDat) {Observed $\hist\tests$};
  \node[latent] (Hid) at (Data |- Hyper) {Future $\hist\tests$};
  \node[latent, right=2em of Data, align=center] (Targeti) {Onset, peak week,\\ peak, duration};
  % xxx (targets...) of $\tests$
  % \node[latent, right of=Data] (Targeti) {$\Target{i}=\targetfun{i}(\Obs,\Hid)$};
  \draw[->,>=stealth] (TrainFit) to (TrainDat);
  \draw[->,>=stealth,dashed,bend right] (TrainDat) to
  node[pos=0.45,left,align=right]{Apply trend \\ filtering} (TrainFit);
  \draw[->,>=stealth] (Hyper) to (TrainFit);
  \draw[->,>=stealth,dashed,bend right] (TrainFit) to
  node[pos=0.45,left,align=center]{Fit uniform \\ distributions} (Hyper);
  \draw[->,>=stealth] (TrainFit) to node[midway,above]{Sample} (Params);
  \draw[->,>=stealth] (Hyper) to node[midway,sloped,above]{Sample} (Params);
  \draw[->,>=stealth] (Params) to (Obs);
  \draw[->,>=stealth,dashed,bend right] (Obs) to
  node[pos=0.45,left=0.2em,align=center]{Weight \\ curve} (Params);
  % todo fix this... maybe "draw nonnoisy curve" on  the forward link, have likelihood as another variable with the parameters and then the back link makes sense?
  \draw[->,>=stealth] (Params) to
  node[pos=0.6,left,align=center]{Draw \\ noisy curve} (Hid);
  \draw[->,ultra thick,>=stealth] (Obs) to
  node[pos=0.35,right,align=right] {Calculate \\ target} (Targeti);
  \draw[->,ultra thick,>=stealth] (Hid) to
  node[pos=0.35,right=0.75em] {Calculate target} (Targeti);
  % Begin from http://tex.stackexchange.com/questions/64505/how-to-make-each-pdf-page-of-a-beamer-output-have-an-opaque-background-when-it-i.
  \begin{pgfonlayer}{background}
    \fill[white]
    (current bounding box.south east)
    rectangle
    (current bounding box.north west);   
  \end{pgfonlayer}
  % End from http://tex.stackexchange.com/questions/64505/how-to-make-each-pdf-page-of-a-beamer-output-have-an-opaque-background-when-it-i.
\end{tikzpicture}
\end{condpreview}
% todo note what weeks are used for conditioning (vs those used for pinning) in the text, potentially make distinction in diagram

  \caption{%
\textbf{Diagram of the prior and posterior model.}
Shaded nodes, unobserved quantities;
shaded dashed nodes, proprietary data;
unshaded nodes, publicly available data;
thin arrows, dependencies;
thick arrows, deterministic dependencies;
textual annotations, descriptions of how we incorporate dependencies.
  }
  \label{fig:framework_net}
\end{condfigure}

\subsubsection{Incorporating non-final and digital data}
At the time that forecasts were generated,
GFT estimates were available for the current week and previous week,
while ILINet wILI measurements were available only for times further in the past.
We produced one set of forecasts using the latest ILINet data by itself,
and another that incorporated GFT data.
We considered two methods of including GFT data:
(i) using GFT estimates only for the two weeks in which ILINet data was not yet available, and
(ii) also using GFT estimates in place of recent ILINet values which may be revised significantly in the future.
Since GFT attempts to minimize RMSE on the logit scale~\cite{Ginsberg2009},
we performed linear regression to reduce the RMSE on the non-logit scale that our framework works with.
% todo check and cite more recent GFT papers?

% Results and Discussion can be combined.
\section{Results}

% We only support three levels of headings, please do not create a heading level below \subsubsection.

\subsection{Predictions for the 2013--2014 season}

For the CDC challenge,
we generated biweekly forecasts
from December 5 (epidemiological week 49) to March 27 (week 9),
for the nation as a whole, and individually for each the 10 HHS regions.
Included below is a summary of our current framework's forecasts throughout the season,
based on revised data and no GFT.
We display 10 draws from the posterior representing likely wILI curves,
as well as the posterior mean and $5$th and $95$th posterior percentiles
for the wILI value for each week.
The posterior provides a histogram for each of the four forecasting targets;
we report the mean target value as a point prediction.

\newcommand\historyplot[1]{%
\begin{condfigure}[p]
  \centering
  \begin{condpreview}
  \includegraphics[width=0.49\linewidth]{history_plots_paper_scale#1forecast.pdf}\hfill
  \includegraphics[width=0.49\linewidth]{history_plots_paper_scale#1peak.pdf}

  \includegraphics[width=0.33\linewidth]{history_plots_paper_scale#1onset.pdf}\hfill
  \includegraphics[width=0.33\linewidth]{history_plots_paper_scale#1peakweek.pdf}\hfill
  \includegraphics[width=0.33\linewidth]{history_plots_paper_scale#1duration.pdf}
  \end{condpreview}
  \pgfmathparse{int(mod(#1+22-1,52)+1)}
  \caption{%
\textbf{2013--2014 national forecast, epidemiological week \pgfmathresult, current framework using final wILI.}
% Forecast (upper left):
% grey, CDC baseline --- the threshold used for onset and duration calculations;
% black circles, data used to make forecast;
% multicolored lines, individual curves from posterior;
% thick black line, pointwise mean of posterior curves;
% dotted blue lines, pointwise $5\%$--$95\%$ credible intervals.
% Peak (top right), onset, peak week, duration (bottom three):
% histogram, distribution of target values in the posterior;
% ``Pt Pred'' (point prediction), the posterior mean target value;
% ``Obs'', the observed value,
% ``Abs Err'', the absolute error of the point prediction.
``Pt Pred'' (point prediction), the posterior mean target value;
``Obs'', the observed value,
``Abs Err'', the absolute error of the point prediction.
}
  \label{fig:w\pgfmathresult}
\end{condfigure}
}
% todo fix figure plot sizes, aspect ratios, legend sizes, etc.
% xxx if include historical range versions, note that onset and duration still use 2 as onset threshold

% todo reference figures in each of these history plot subsubsections
\subsubsection{Week 49 (December 5) forecast, using wILI data through week 47}
During the week of the first forecast,
all of the available wILI values are below the CDC onset threshold.
Predictions for the onset, as shown in \autoref{fig:w49},
are concentrated near the actual value,
and the error in the point prediction is fairly small.
Much of this error can be attributed to the sudden jump in wILI at the onset,
which corresponds to Thanksgiving week.
The number of patients seen per reporting provider in ILINet drops noticeably
on Thanksgiving week and even more significantly during winter holidays;
at these times, there is a systematic bias towards higher wILI values.
The forecasts for the overall wILI curves and the other three targets contain much more uncertainty,
as shown by wider histograms that more closely resemble the prior distribution.
The peak of the epidemic could potentially occur early or late, and be mild or strong.
  \historyplot{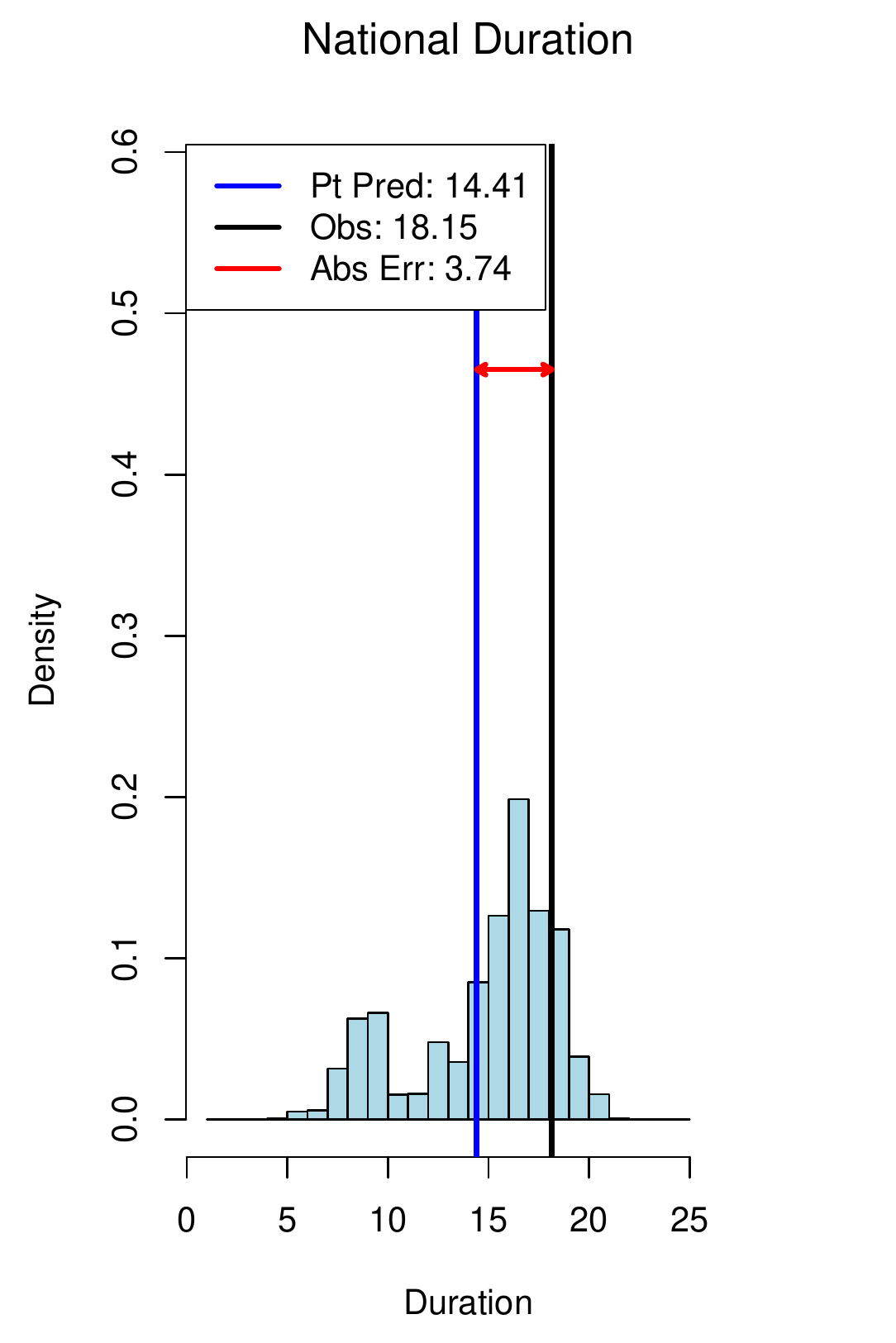}
% \item Week 51 (29)
%   \historyplot{29}
\subsubsection{Week 1 (January 2) forecast, using wILI data through week 51}
\autoref{fig:w1} shows that,
with data available up to the week before the sudden peak,
the framework matches the observed wILI trajectory 
fairly closely with many of the posterior draws.
The sudden peak can be explained as a combination of elevated ILI-related visits combined
with a relative decrease in unrelated visits associated with winter holidays.
The framework selects
posterior curves with slightly later peaks of similar height, as well as
seasons with much later peaks, which contain secondary peaks around the winter holidays.
The onset has already been confirmed,
so its histogram is a point mass.
Duration predictions narrow around the actual duration.
  \historyplot{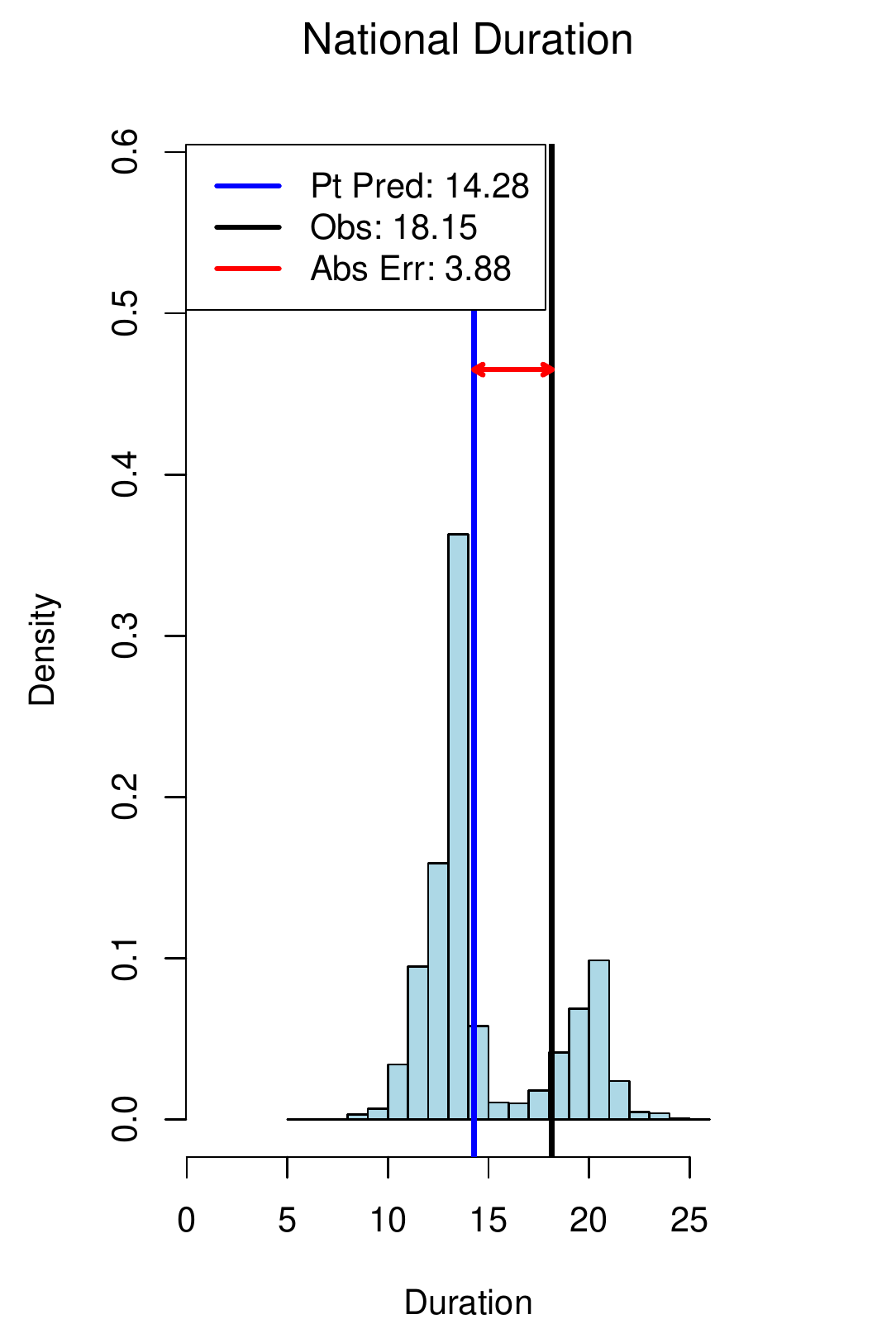}
\subsubsection{Week 3 (January 16) forecast, using wILI data through week 1}
\autoref{fig:w3} shows that,
after the sudden peak, the posterior for the 2013--2014 epidemic
contained primarily transformed versions of the 2006--2007 curve, 
which featured a relatively large secondary peak around winter holidays,
followed by a primary peak in early February.
This tendency to ``latch'' onto a particular shape
is one of the reasons why transformed versions of curves were used
instead of just the curves themselves.
Given the limited amount of historical wILI data available,
including transformations provides a larger number of reasonable curves from which to pick.
However, the latching phenomenon still occurs, and degrades performance in this forecast.
  \historyplot{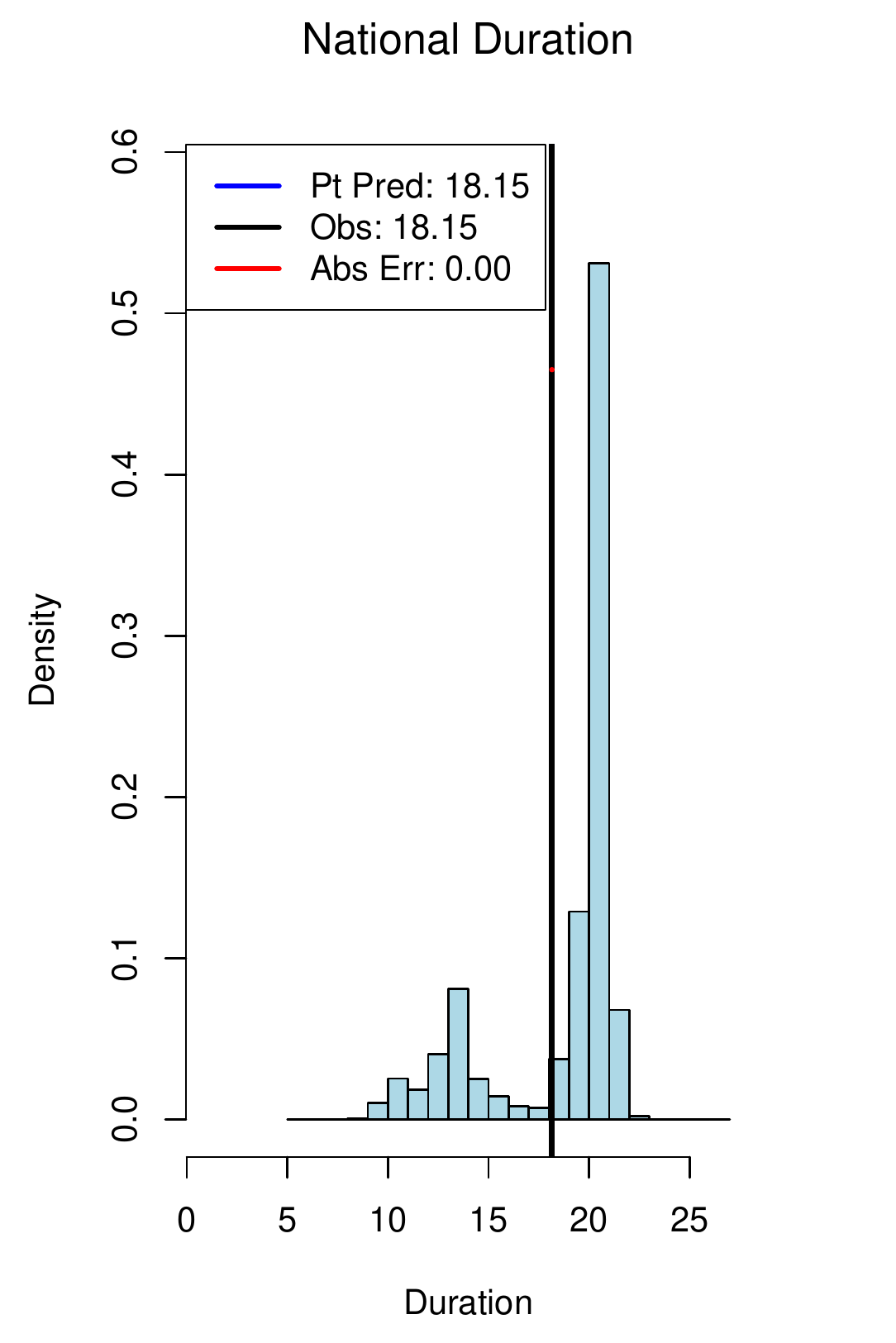}
Subsequent forecasts continue to predict another, later, primary or secondary peak,
until some time late in the season, in which forecasts match the falling tail of the epidemic curve.
% \item Week 9 (39)
%   \historyplot{39}

\ref{sec:full_forecast_history} contains forecasts like those referenced above for the entire 2013--2014 season.

\subsection{Point prediction trends}
\autoref{fig:trends} shows the observed forecasting target values for the 2013--2014 season,
and predictions over time for the current framework,
older versions used to generate submissions with different transformations and curve weighting methods,
and Epicast, a system used to collect and aggregate human-generated predictions.
% todo should Epicast be brought into the picture in this paper?
% todo include some other baseline(s) in this graph?
  \begin{condfigure}[p]
    \centering
    \begin{condpreview}
    \includegraphics[width=0.49\linewidth]{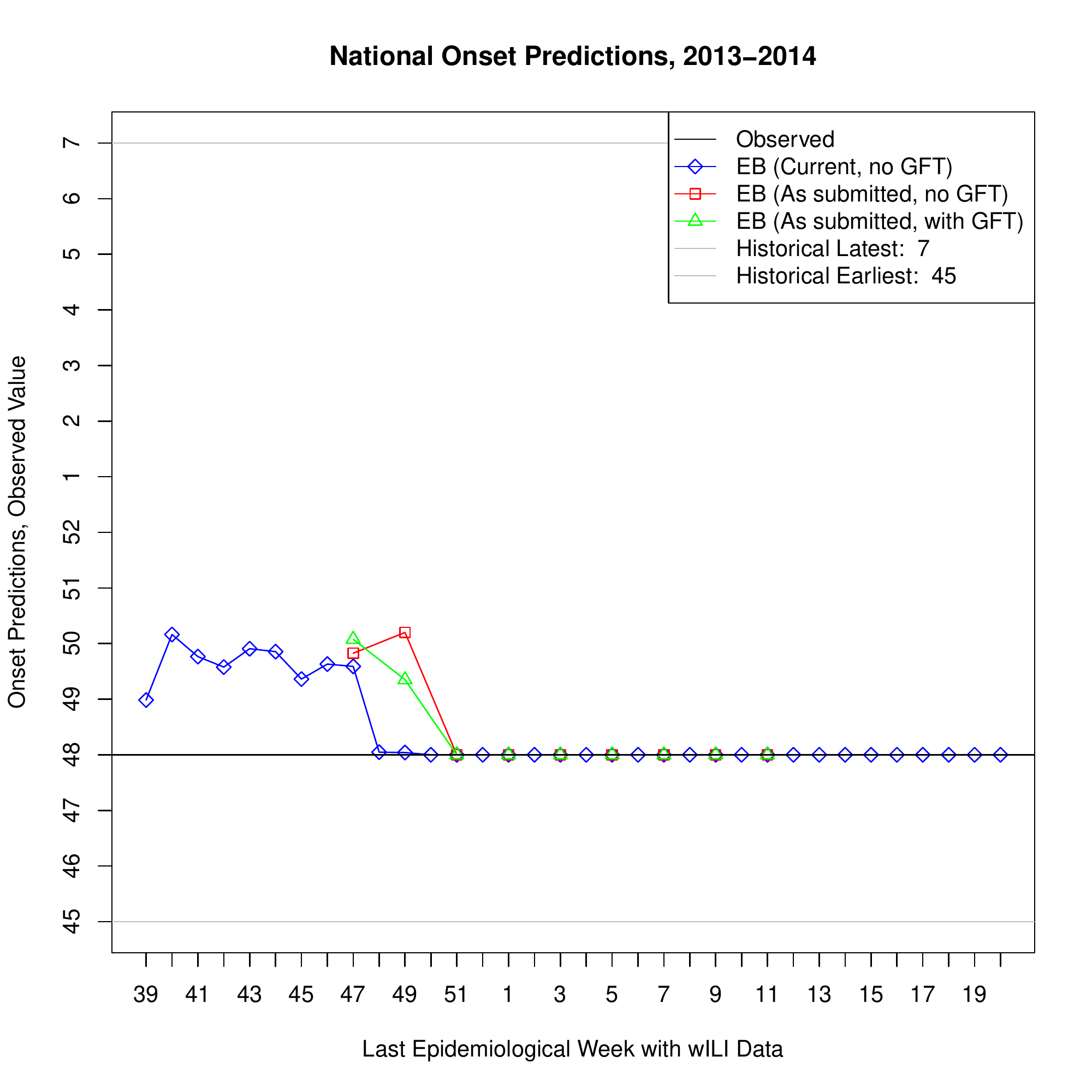}
    \includegraphics[width=0.49\linewidth]{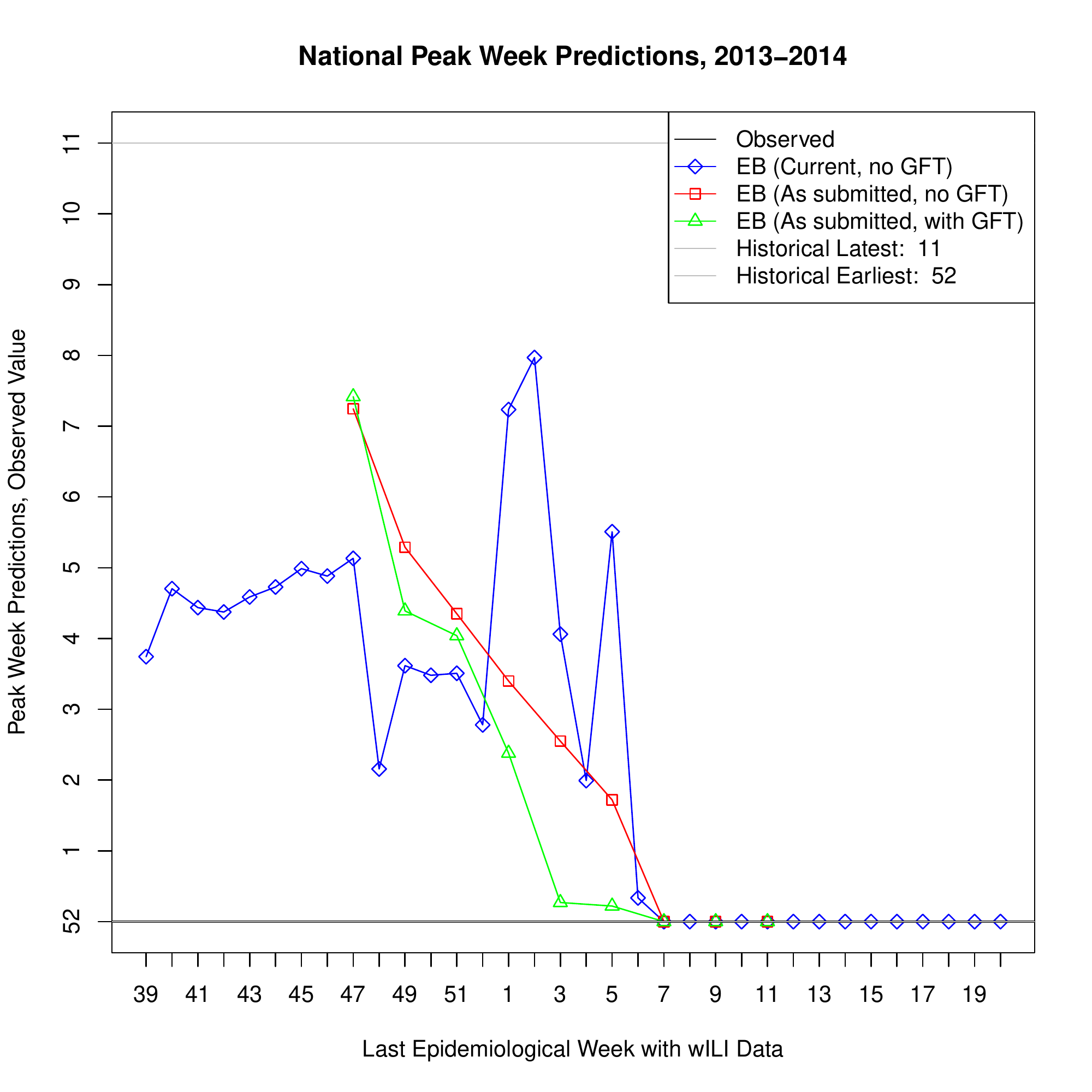}

    \includegraphics[width=0.49\linewidth]{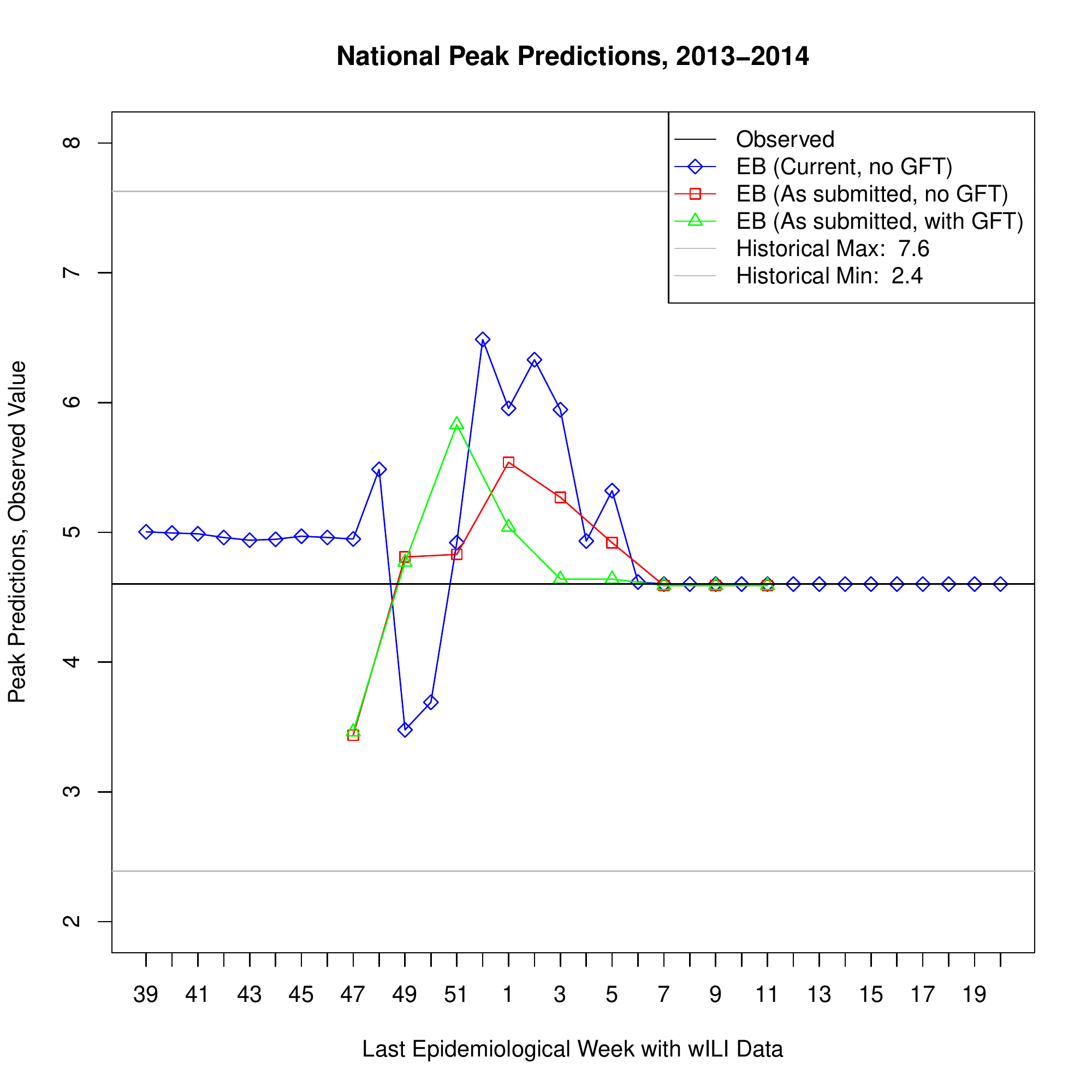}
    % todo fix historical max/min showing too many digits
    \includegraphics[width=0.49\linewidth]{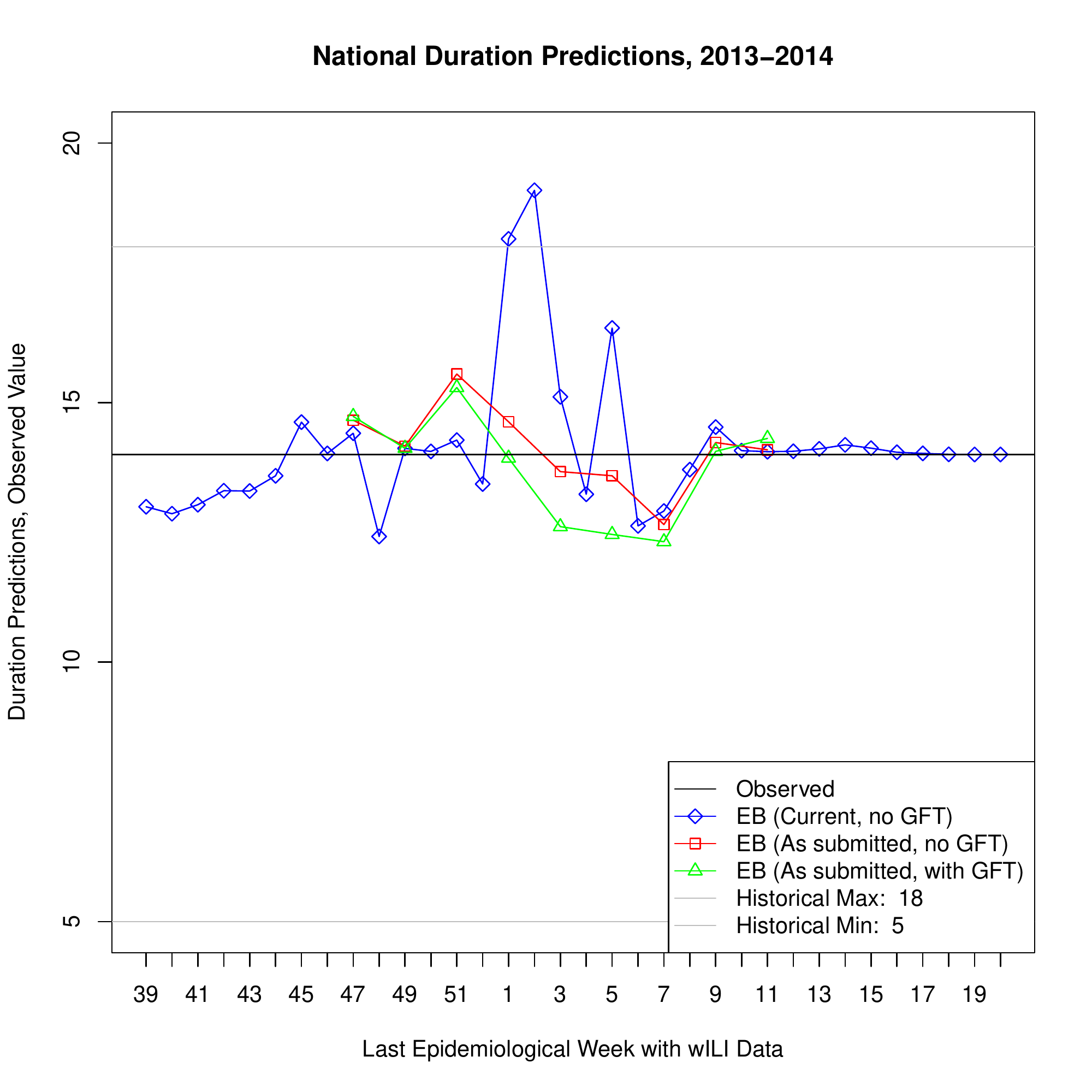}
    \end{condpreview}

%     \caption{Observed forecasting target values for the 2013--2014 season,
% and predictions over time for the current framework, older versions used to generate submissions with different transformations and curve weighting, and an aggregate of human-generated predictions}
    \caption{\textbf{Point predictions and observed values of the forecasting targets for the 2013--2014 season.}  Black, observed target value; blue, our current framework's predictions using revised ILINet wILI data; red, our submitted point predictions using ILINet data only; green, our submitted point predictions that used both ILINet and GFT data; dark goldenrod, the mean of human forecasts collected with Epicast on revised wILI data.  Historical target value ranges exclude the 2009--2010 and 2013--2014 seasons.}
    \label{fig:trends}
  \end{condfigure}
The small error in the onset before it occurred,
as well as the underestimation of the peak in weeks 49 and 50, 
can be attributed to not factoring in holiday effects;
at least some of these effects are smoothed out by the trend filtering process,
or shifted to different times and heights by the peak week and height transformations.
Later errors in the peak week, peak, and duration result from latching onto
transformed versions of one or two past epidemic curves with two peaks.
Including additional curves in the prior,
improving selection of transformations,
accounting for holiday effects,
and incorporating additional types of data
may help increase performance in this case.
However, when considering these types of changes,
we must be careful not to select modifications
only to give good performance only on the 2013--2014 season;
that would be fitting, rather than prediction.
The ideal way to compare different approaches would be to
analyze the error of true, prospective forecasts over many seasons;
however, this comes at the prohibitive cost of waiting through these seasons.
Instead, we prefer to use cross-validated error when predicting past seasons
rather than performance on a single season.

\subsection{Estimated average error from cross-validation}
We used leave-one-out cross-validation to approximate
the average error in the framework's forecasting target point predictions
as a function of the epidemiological week of the last wILI observation used.
For each historical season $\cvtests$,
we produced forecasts
using the rest of the historical seasons
to build the prior,
and recorded the average error of our point predictions across these 15 seasons
for each week in the flu season.
While this approach allows later seasons to be used
while predicting earlier seasons,
this should not introduce too much bias
because the framework treats the different seasons as if they were independent.
The advantage of including these later seasons
is that each one uses $14$ epidemic curves
to build its prior,
and should give a better idea of what
to expect when making forecasts with
a similar number of ``past'' seasons.
Another detail to note is that these error estimates
were generated using the final revision of the wILI data, and
do not include any effects from
approximating the most recent wILI values from
the tentative values available in real-time.
% leave-one-out, use fut seasons when pred'ing same season (Viboud did sameish)

Figures~\ref{fig:cv_onset}, \ref{fig:cv_peakweek}, \ref{fig:cv_peak}, and~\ref{fig:cv_duration}
show the cross-validated error for our current empirical Bayes framework,
as well a few other approaches,
for each for the four forecasting targets.
% todo describe approaches, ...
The methods for predicting $\tarj(\hist\cvtests)$ are summarized below.
\begin{description}
\item[Baseline (Mean of Other Seasons):]
takes the average target value across the 14 other seasons,
completely ignoring any data from the current season.
\item[Pinned Baseline (Mean of Other Seasons, Conditioned on Current Season to Date):]
constructs 14 possible wILI trajectories for the current season
by using the available observations for previous weeks
and other historical curves for future weeks;
reports the mean target value across these 14 trajectories.
\item[Pointwise Percentile~\cite{van2014risk}:]
Constructs a single possible future wILI trajectory
using the pointwise $q$th quantile from other seasons;
estimates an appropriate value of $q$ from the observed data so far.
\item[$k$ Nearest Neighbors:]
Uses a method similar to existing systems for shorter-term prediction~\cite{viboud2003prediction}
to identify $k$ sections of other seasons' data that best match recent observations,
and uses them to construct and weight $k$ possible future wILI trajectories.
\item[Empirical Bayes (Transformed Versions of Other Seasons' Curves):]
Our current framework, using transformed versions of other seasons' curves to form the prior.
\item[Empirical Bayes (SIR Curves):]
Our current framework, using scaled and shifted SIR curves rather than other seasons' curves to form the prior.
% todo fix tense problems...
% todo fix poor descriptions
\end{description}

From the graphs, we see that the empirical Bayes framework
gives competitive or lower mean absolute error
in forecasting target point predictions
than the best of the other baselines
consistently across many prediction weeks.
An important advantage of this approach
over many of the other baselines
is that it provides a smooth distribution
over possible curves and target values,
rather than just a single point.
% xxx (((even though the smoothness is sort of artificial and we can't really trust the distribution)))
From this distribution,
we can calculate point predictions
to minimize some expected type of error or loss,
build credible intervals,
and make probabilistic statements
about future wILI and target values.

\begin{condfigure}[p]
  \centering
  \begin{condpreview}
  \includegraphics[width=\linewidth]{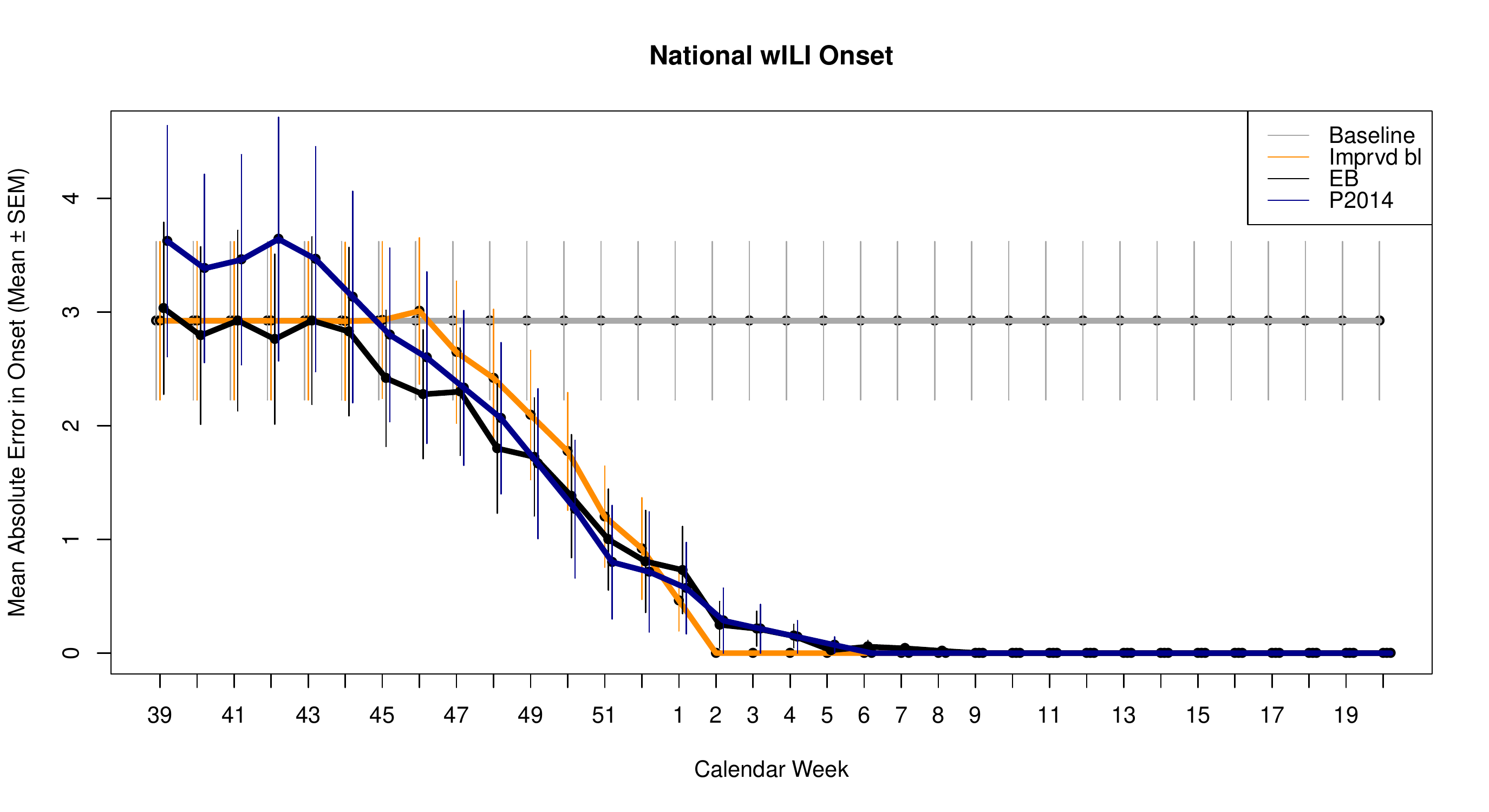}
  \vspace{-1.5\baselineskip}
  \end{condpreview}
  \caption{%
\textbf{Cross-validated mean absolute error estimates for onset point predictions.}
(The onset was defined based on the 2\% national threshold set by CDC for the 2013--2014 season.)
}
  \label{fig:cv_onset}
\end{condfigure}
\begin{condfigure}[p]
  \centering
  \begin{condpreview}
  \includegraphics[width=\linewidth]{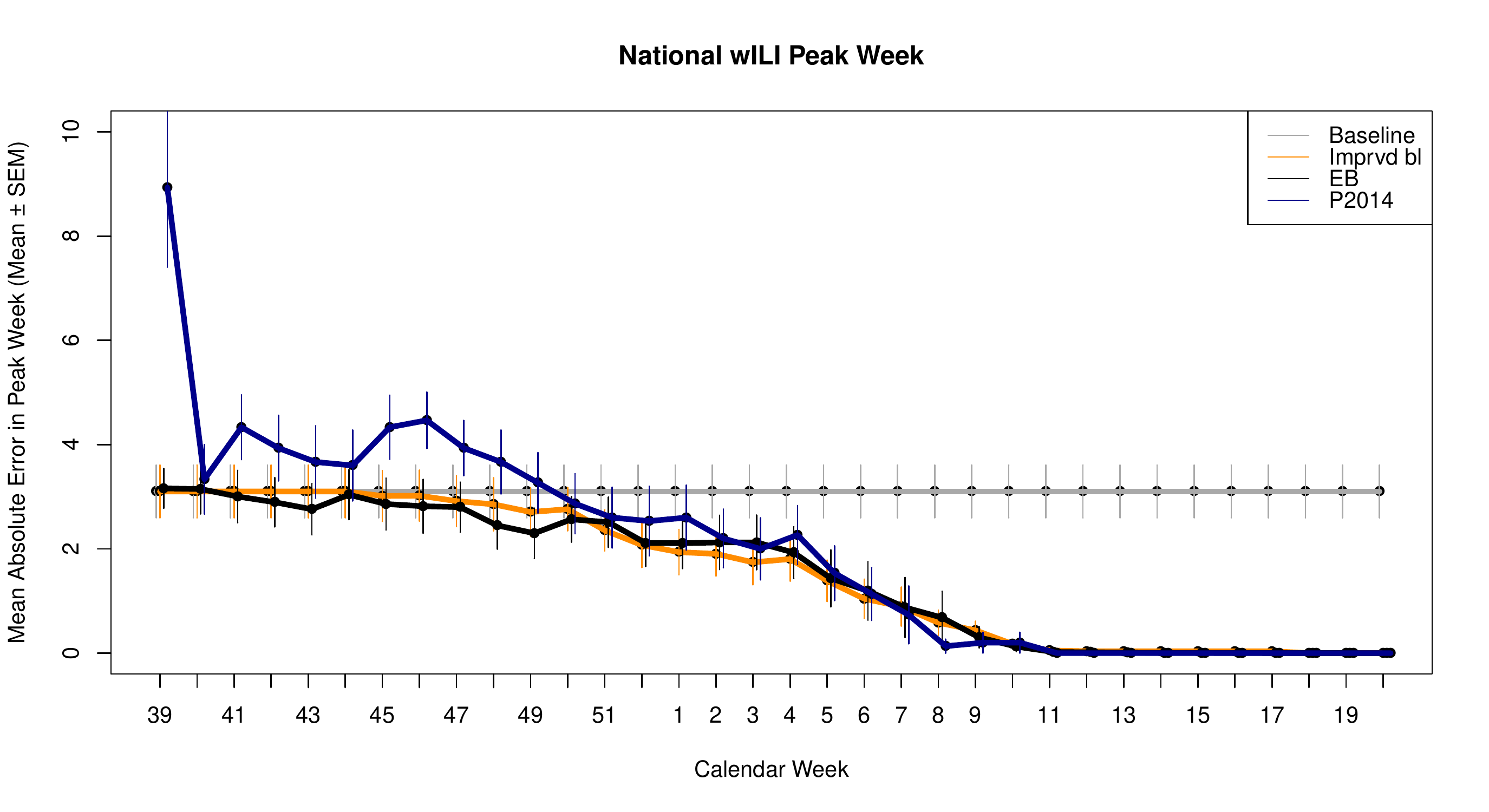}
  \vspace{-1.5\baselineskip}
  \end{condpreview}
  \caption{%
\textbf{Cross-validated mean absolute error estimates for peak week point predictions.}
}
  \label{fig:cv_peakweek}
\end{condfigure}
\begin{condfigure}[p]
  \centering
  \begin{condpreview}
  \includegraphics[width=\linewidth]{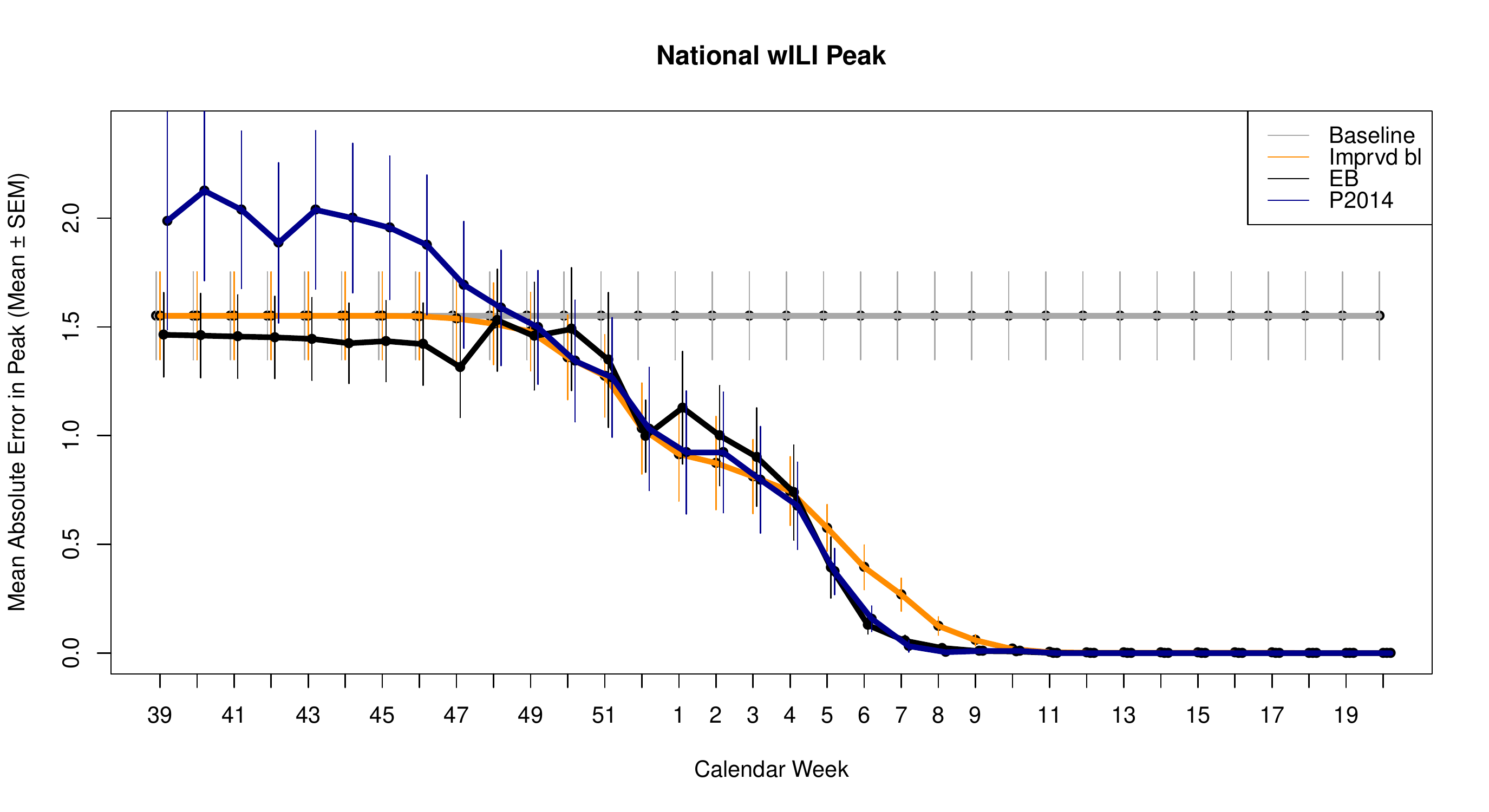}
  \vspace{-1.5\baselineskip}
  \end{condpreview}
  \caption{%
\textbf{Cross-validated mean absolute error estimates for peak height point predictions.}
}
  \label{fig:cv_peak}
\end{condfigure}
\begin{condfigure}[p]
  \centering
  \begin{condpreview}
  \includegraphics[width=\linewidth]{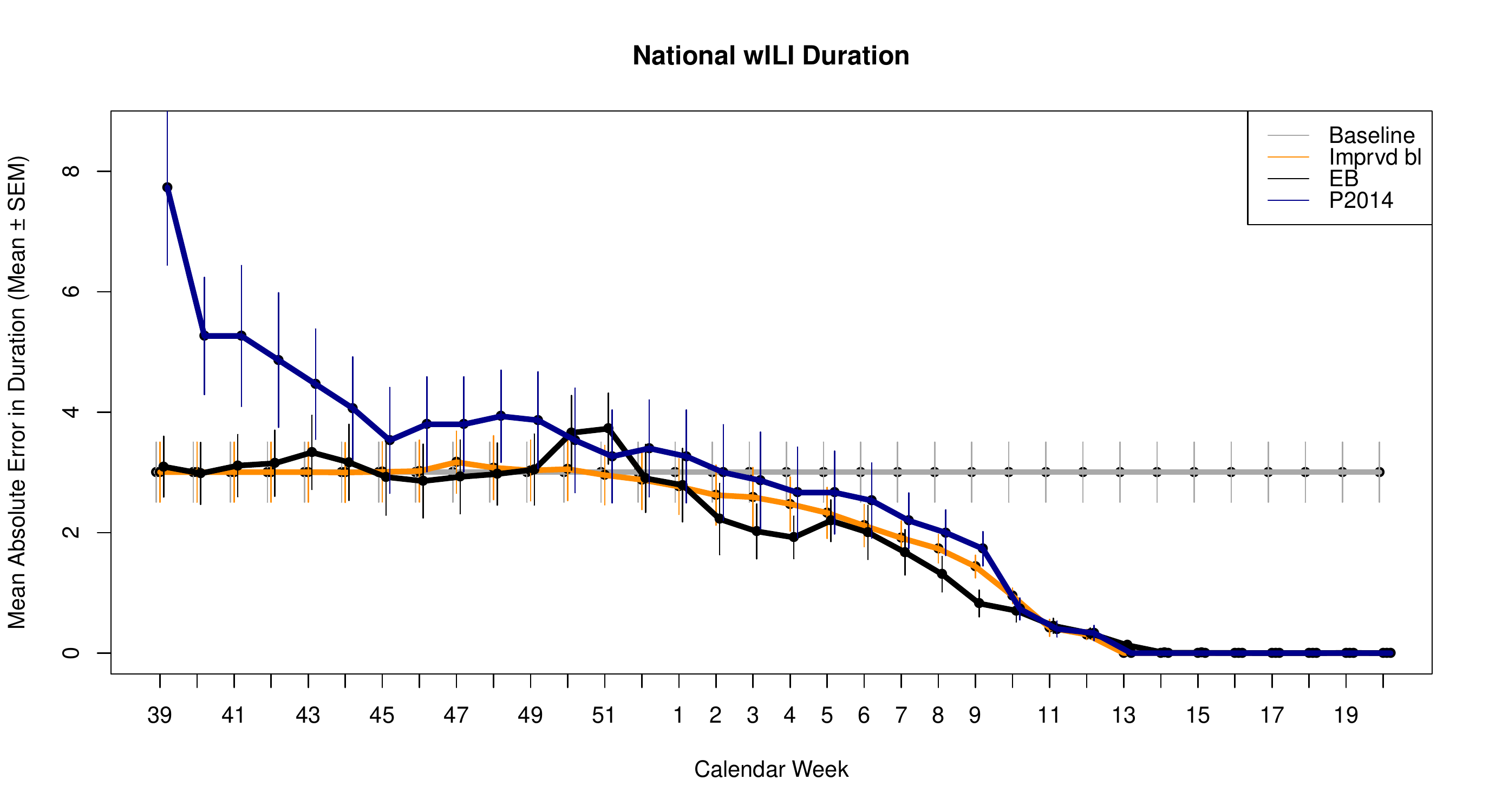}
  \vspace{-1.5\baselineskip}
  \end{condpreview}
  \caption{%
\textbf{Cross-validated mean absolute error estimates for duration point predictions.}
(The duration was defined based on the 2\% national threshold set by CDC for the 2013--2014 season.)
}
  \label{fig:cv_duration}
\end{condfigure}
% todo white background in figures, fix plot and legend sizes

\section{Discussion}

We developed an empirical Bayes approach to forecasting epidemic curves and targets,
and applied it to wILI estimates to generate predictions
for the 2013--2014 influenza season as part of a CDC challenge.
Our method's forecasts for the season were reasonable to the human eye,
and cross-validated error estimates indicate
that it competes with or improves upon results
from various baseline predictors.
This method generates a distribution over future wILI curves and forecasting targets,
rather than just point predictions.
The framework does not require or rely on mechanistic models,
which often will not match well with observed data,
but instead generates possible epidemic curves
from past history with a few reasonable transformations.

A potential downside is that
we also gain no insight into the process underlying the epidemic.
Since it is nonmechanistic, our framework can be easily applied to other epidemic curves.
We have already used it to predict dengue incidence in the 2014 World Cup game cities
with little modification~\cite{van2014risk},
and expect that application to additional diseases will require little adjustment,
and could be considered as a baseline for other, more specialized, predictors.
% todo cite dengue paper
It should be possible to make important decisions,
such as what transformations to use,
automatically by minimizing cross-validated prediction error on historical data.

We are investigating many ways to improve our approach's performance.
Better ways of incorporating non-final wILI and other surveillance data,
complemented by short-term predictors such as regression,
should improve whole-season predictions as well.
Our current framework only uses ILINet and GFT,
but can include additional sources of data
such as Twitter activity, thermometer sales, and lab testing data.
Modeling and adjusting for holiday effects
in each data source may also improve accuracy.
For now, we have treated regional epidemics as independent,
but have found spatial correlations in historical data;
shrinking forecasts together based on proximity may improve our results.
% todo make less awkward discussion section / end

\section{Acknowledgments}

We thank Matt Biggerstaff, Lyn Finelli, and the CDC
for organizing the challenge and the followup workshop,
and for helpful discussions regarding influenza surveillance in the U.S.

% MIDAS grant:
Research reported in this publication was supported by the National
Institute Of General Medical Sciences of the National Institutes of
Health under Award Number U54 GM088491. The content is solely the
responsibility of the authors and does not necessarily represent the
official views of the National Institutes of Health.
% LB NSF GRF:
This material is based upon work supported by the National Science
Foundation Graduate Research Fellowship Program under Grant
No. DGE-1252522.
% The Fellow is responsible for assuring that every publication of material (including World Wide
% Web pages) based on or developed during the fellowship, except scientific articles or papers
% appearing in scientific, technical or professional journals, contains the following disclaimer:
Any opinions, findings, and conclusions or recommendations expressed in this material
are those of the authors and do not necessarily reflect the views of the National Science
Foundation.
% DF:
DF was a predoctoral trainee supported by NIH T32 training grant T32
EB009403 as part of the HHMI-NIBIB Interfaces Initiative.

\bibliography{Epi-Forecasting,eb_cdc_chall_1314_addit}

\setcounter{table}{0}%
\renewcommand{\thetable}{S\arabic{table}}%
\setcounter{figure}{0}%
\renewcommand{\thefigure}{S\arabic{figure}}%
\section{Supplementary Information}

\subsection{Full Forecast History, 2013--2014 Season}\label{sec:full_forecast_history}

Figures~\ref{fig:fullw41}--\ref{fig:fullw22} show the full forecast history for the 2013--2014 flu season,
using our latest framework and the final revision of the wILI values.

\renewcommand\historyplot[1]{%
\begin{condfigure}[H]
  \begin{condpreview}
  \centering
  \includegraphics[width=0.49\linewidth]{history_plots#1forecast.pdf}\hfill
  \includegraphics[width=0.49\linewidth]{history_plots#1peak.pdf}

  \includegraphics[width=0.33\linewidth]{history_plots#1onset.pdf}\hfill
  \includegraphics[width=0.33\linewidth]{history_plots#1peakweek.pdf}\hfill
  \includegraphics[width=0.33\linewidth]{history_plots#1duration.pdf}
  \end{condpreview}
  \pgfmathparse{int(mod(#1+22-1,52)+1)}
  \caption{%
\textbf{2013--2014 national forecast, Week \pgfmathresult, using the current framework and the final wILI values.}
The wILI observations for the week of the forecast and preceding week, are not used in the forecast.
% Forecast (upper left):
% grey, CDC baseline --- the threshold used for onset and duration calculations;
% black circles, data used to make forecast;
% multicolored lines, individual curves from posterior;
% thick black line, pointwise mean of posterior curves;
% dotted blue lines, pointwise $5\%$--$95\%$ credible intervals.
% Peak (top right), onset, peak week, duration (bottom three):
% histogram, distribution of target values in the posterior;
% ``Pt Pred'' (point prediction), the posterior mean target value;
% ``Obs'', the observed value,
% ``Abs Err'', the absolute error of the point prediction.
``Pt Pred'' (point prediction), the posterior mean target value;
``Obs'', the observed value,
``Abs Err'', the absolute error of the point prediction.
}
  \label{fig:fullw\pgfmathresult}
\end{condfigure}
}
\foreach \n in {19,...,52}{
  \historyplot{\n}
}

\end{document}